\documentclass[]{spie}  

\usepackage{amsmath,amsfonts,amssymb}
\usepackage{graphicx}
\usepackage[colorlinks=true, allcolors=blue]{hyperref}

\usepackage{setspace}
\usepackage{tocloft}
\usepackage[utf8]{inputenc}
\usepackage{rotating}
\usepackage{array}
\usepackage{lineno}

\title{Jitter Sensing and Control for Multi-Plane Phase Retrieval}

\author[a]{Caleb G. Abbott}
\author[a]{Justin R. Crepp}
\author[a]{Brian Sands}
\affil[a]{University of Notre Dame, Physics and Astronomy, Notre Dame, IN 46556, USA}
\authorinfo{Further author information: (Send correspondence to C.G.A.)\\C.G.A.: E-mail: cabbott3@nd.edu}

\begin{document} 
\maketitle

\begin{abstract}
The family of multi-plane phase retrieval sensors, such as the curvature and nonlinear curvature wavefront sensors (WFS), contain tip/tilt information embedded in their signals. We have built a nonlinear curvature WFS to study different wavefront reconstruction methods and test the ability to extract tip/tilt information. Using reliable and fast centroiding algorithms, combined with knowledge of the measured $z$-distance to each measurement plane, we demonstrate that image jitter may be sensed and compensated for using a fast steering mirror and the WFS in closed loop. This approach obviates the need for peripheral components such as quad-cells or access to a separate scientific imaging channel. Our laboratory experiments validate tip/tilt estimation and correction using nlCWFS data, achieving tip/tilt accuracy of $\pm0.1\,\lambda/D$ for an unaberrated beam and better than $\sim \pm0.5\,\lambda/D$ in the presence of aberrations, consistent with prior numerical simulations. We further demonstrate a closed-loop tip/tilt control implementation and show a qualitative improvement in the stability and overall quality of multi-plane phase retrieval reconstructions.
\end{abstract}


\keywords{wavefront sensing, adaptive optics, wavefront reconstruction algorithms}

\section{INTRODUCTION}\label{sec:intro}

Atmospheric turbulence contains most of its power in the lowest spatial frequencies, manifesting as tip and tilt aberrations when light is collected by a ground-based telescope. Left uncorrected, these low-order modes spatially smear or blur images that would otherwise be diffraction-limited; in the case of spectroscopy, significant signal may be lost when the target drifts beyond the limits of a slit or fiber \cite{Bechter2020,Crepp2014}. Residual tip/tilt errors can also degrade the closed-loop reconstruction accuracy of higher-order aberrations of Adaptive Optics (AO) systems\cite{Tyson2022}. With multi-plane diffractive WFS, such as the nonlinear Curvature Wavefront Sensor (nlCWFS), Abbott et al. 2025 have shown that accurate recovery of high-order spatial modes depends critically on precise tip/tilt sensing and jitter control \cite{Abbott2025}.

The nlCWFS offers excellent sensitivity and resilience to scintillation compared with other WFS such as the Shack-Hartmann Wavefront Sensor (SHWFS) \cite{Guyon2010,Mateen2011,Crass2014b,Crepp2020,Letchev2024}. Typically four measurement planes are employed, each offset from the pupil along the optical axis, and a phase retrieval algorithm is used to reconstruct the wavefront phase and amplitude. Owing to the adaptable placement of $z$-distances at each measurement plane, this architecture provides design flexibility for challenging applications with both monochromatic and broadband light \cite{Mateen2011,Letchev2022,Potier2023,Ahn2023b}.

Jitter sensing can be implemented in the nlCWFS using geometric optics and/or wave optics, enabling tip/tilt correction without ancillary devices and allowing more light to reach scientific instruments\cite{Abbott2025}. By analyzing spatial offsets in each measurement plane, tip/tilt modes may be traced along the optical path, enabling extraction of their direction and magnitude. A multi-plane configuration provides redundancy and cross-validation as individual planes have tip/tilt information encoded in the image. Small errors in tip/tilt estimation can further cascade into large uncertainties in reconstruction of the overall wavefront, so establishing a reliable and accurate method for extracting tip/tilt is of high priority. Precise jitter control also leads to improved phase unwrapping performance \cite{Huerta2025}. 

Conventional tip/tilt sensing and correction typically relies on dedicated detectors in a separate optical path, such as quad-cell or focal-plane tip/tilt sensors feeding a fast steering mirror, as implemented in many astronomical AO systems \cite{Avicola1998}. In laser guide star AO, separate tip/tilt cameras are also commonly used (often at near-infrared wavelengths) to improve sky coverage and sensitivity \cite{Wizinowich2014}. In contrast, Differential Image Motion Monitors (DIMMs) are widely used for site testing and seeing characterization by measuring differential image motion between sub-apertures \cite{Sarazin1990,Tokovinin2007}. While effective, dedicated tip/tilt sensors generally require additional hardware and beam-splitting optics, increasing alignment complexity and reducing photon throughput. The approach demonstrated here avoids ancillary detectors by extracting tip/tilt directly from the nlCWFS intensity measurements already acquired for phase reconstruction.

While this work focuses on laboratory validation, the technique has potential applications in systems that require low-order wavefront sensing and jitter correction. These include, but are not limited to: ground-based astronomical telescopes, where photon throughput is critical; and free-space optical communication systems, which are sensitive to pointing stability. In its current implementation, the method is best suited for correcting low-frequency tip/tilt errors, such as those arising from mechanical vibrations or tracking jitter. The ability to recover tip/tilt directly from the wavefront sensing data offers a path to simplifying optical systems while preserving or improving performance, particularly in photon-limited environments.  

Abbott et al. (2025) \cite{Abbott2025} performed simulations evaluating the precision, accuracy, and latency of multiple centroiding methods for the nlCWFS under a range of conditions. By examining residual tip/tilt, they identified two prospective approaches for centroid estimation: a geometric intensity Weighted Average (WA) method (similarly Weighted Center of Gravity (WCoG)\cite{Nicolle2004}) and physical-optics-based Speckle Pattern Method (SPM). While SPM proved to be most accurate, achieving better than $0.1 \;\lambda/D$ accuracy (more than three times better than WA for an aberrated beam), it was also significantly more expensive computationally, more than 500$\times$ slower. Based on this trade-off, Abbott et al. 2025 recommended the use of fast, empirical algorithms such as WA for real-time applications. The choice of centroiding method thus involves a balance between speed and accuracy, and must ultimately be validated through laboratory experimentation. The objective of this work is to confirm the simulation results reported by Abbott et al. 2025 and demonstrate that an empirical method like WA can indeed achieve diffraction-limited performance as claimed.

To evaluate the simulation results, we have developed a laboratory experiment to compare centroiding precision and accuracy with simulations. Section \ref{sec:exp} describes the optical layout, experimental setup, and algorithms used to test tip/tilt analysis and jitter control. Section \ref{sec:results} presents the outcomes of these tests using the intensity weighted average (WA) method and assessing phase reconstruction quality. Finally, Section \ref{sec:summary} provides conclusions and a discussion of future directions.
\section{EXPERIMENTS AND METHODS}\label{sec:exp}

We aim to substantiate the results from Abbott et al, 2025\cite{Abbott2025} by performing tip/tilt measurements in a lab setting using real hardware in the presence of noise. Experiments involved quantifying centroiding and tip/tilt extraction for unaberrated and aberrated beams. Known amounts of tip/tilt were introduced using a fast steering mirror (FSM) to provide truth measurements for which to compare results.

\subsection{Experimental Design} \label{sec:inst_des}

\begin{figure}
    \centering
    \includegraphics[width=0.9\linewidth]{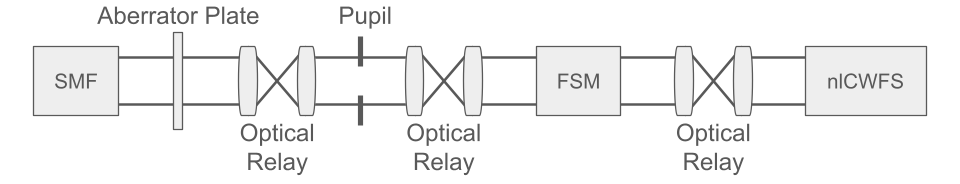}
    \caption{Diagram representing the light path of the experiment. An aberrator plate may be translated through the beam to induce aberrations.}
    \label{fig:system}
\end{figure}

\begin{figure}
    \centering
    \includegraphics[width=0.26\linewidth]{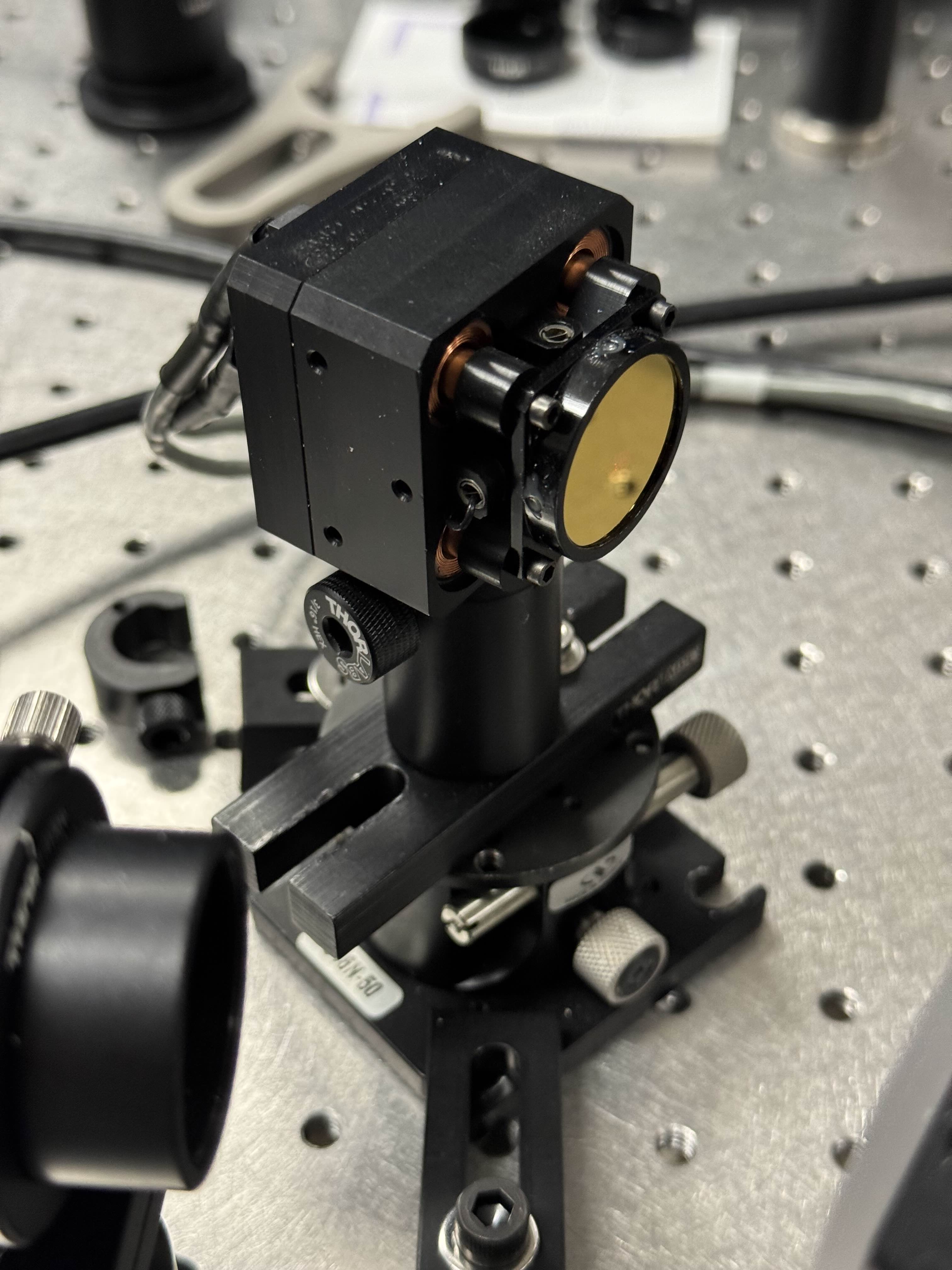}
    \includegraphics[width=0.26\linewidth]{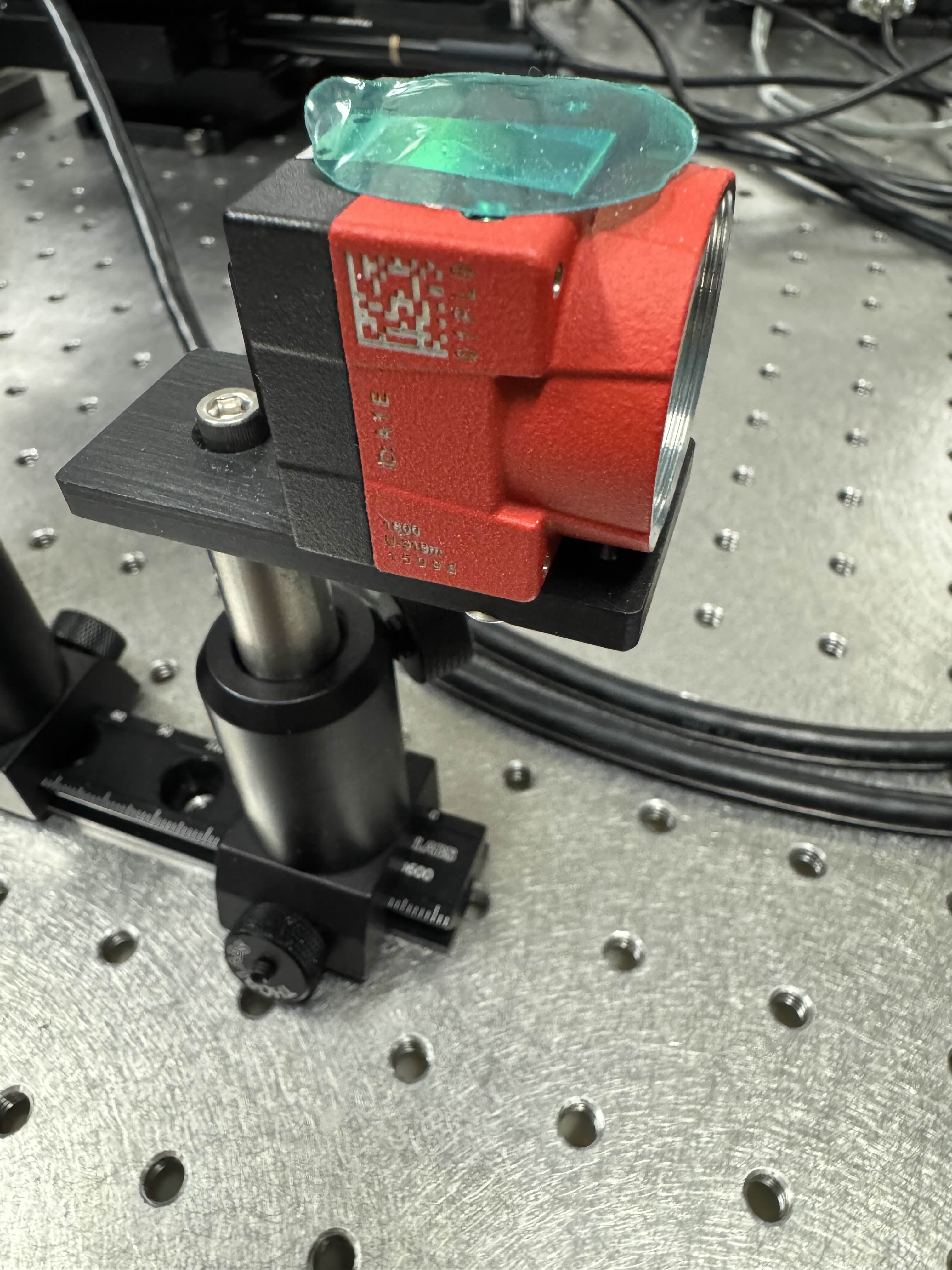}
    \caption{Images of the FSM (left) and tip/tilt sensing camera (right).}
    \label{fig:hardware}
\end{figure}

A diagram of the experimental setup is illustrated in Figure \ref{fig:system}. Collimated light is injected into the system from a single-mode fiber (SMF) illuminated by a HeNe laser operating at $\lambda = 633$ nm. The beam diameter and pupil are defined by a 1 mm pinhole. An aberrating waveplate can optionally be inserted upstream of the pupil to introduce aberrations and scintillation effects. The aberrator plate provides a repeatable, controlled phase disturbance used to evaluate the robustness of centroiding, tip/tilt retrieval, and reconstruction behavior.
 The beam is redirected by a FSM, which is used either to correct tip/tilt offsets for control purposes or introduce known tip/tilt signals for calibration. An optional deformable mirror (DM) is included in the path for closed-loop experiments.\footnote{Results for closed-loop control will be reported in a follow-on article.} After the DM, the beam passes through a 2:1 beam reducer, decreasing its diameter to 0.5 mm to induce diffraction over short distances. At this scale, the $z$-distances for the four nlCWFS imaging planes are set to $\pm 1$ cm (near/inner planes) and $\pm 5$ cm (far/outer planes) as measured from the pupil\cite{Letchev2023}. Images are recorded using an Allied Vision camera with a plate scale of $3.45\; \mu$m. Table~\ref{tab:components} lists components in the experiment. Images of key hardware devices are shown in Figure~\ref{fig:hardware}.

\begin{table}[h]
    \centering
    \begin{tabular}{c|ccc}
    \hline
    \hline
     Component      &  Vendor & Model & Notes \\
     \hline
      HeNe Laser    &   Melles Griot      &   ---   & $\lambda = 633$ nm \\ 
      FSM           &  Optics in Motion         &  OIM5001      & $\pm1.5^{\circ}$ range \\
      DM            &  Boston Micromachines     &  Multi-C-DM     & 140 actuators \\
      CCD           &  Alvium                   &  1800 U-319m  & $2064 \times 1544$ pixels \\
      Aberrator Plate & University of California, Santa Cruz & --- &$r_0: 330-684$ um \\
     \hline
     \hline
    \end{tabular}
    \caption{Experimental components list.}
    \label{tab:components}
\end{table}

\subsection{Centroiding and Calibration} \label{sec:centr_calib}

Reliable tip/tilt extraction depends on accurate centroid estimation and image centering. The primary centroiding method used in this study is the WA method, selected for its balance of speed and accuracy \cite{Thomas2006}. As reported in Abbott et al. (2025) \cite{Abbott2025}, WA achieved approximately $\pm0.3\,\lambda/D$ accuracy in 5.7 ms for a $256 \times 256$ array. The centroid location for each image plane ($n = 1$–$4$) is computed as follows:
\begin{equation}
    x_n = \frac{\sum x_i\, I_n(x_i, y_j)}{\sum I_n(x_i, y_j)}, \quad
    y_n = \frac{\sum y_j\, I_n(x_i, y_j)}{\sum I_n(x_i, y_j)},
\end{equation}
where the summation runs over all pixel coordinates $(i, j)$. Although efficient, WA is known to be sensitive to background noise and intensity outliers\cite{Kong2017}. To evaluate its robustness under realistic conditions, our experiments include both unaberrated and distorted beams. WA centroiding implicitly assumes a compact, approximately symmetric intensity distribution. As asymmetry increases, intensity outliers and nonuniform structure can bias the weighted sum, increasing both centroid error and variability. As tip/tilt correction recenters and stabilizes the intensity patterns, the WA estimate becomes less biased and more repeatable.

\begin{figure}
    \centering
    \includegraphics[width=0.9\linewidth]{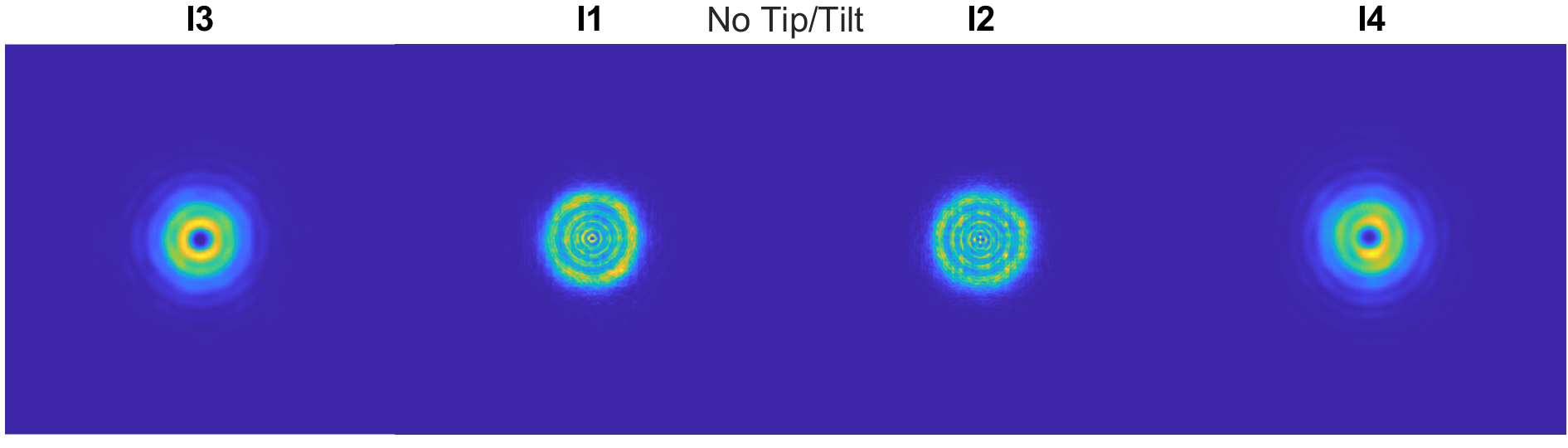}
    \includegraphics[width=0.9\linewidth]{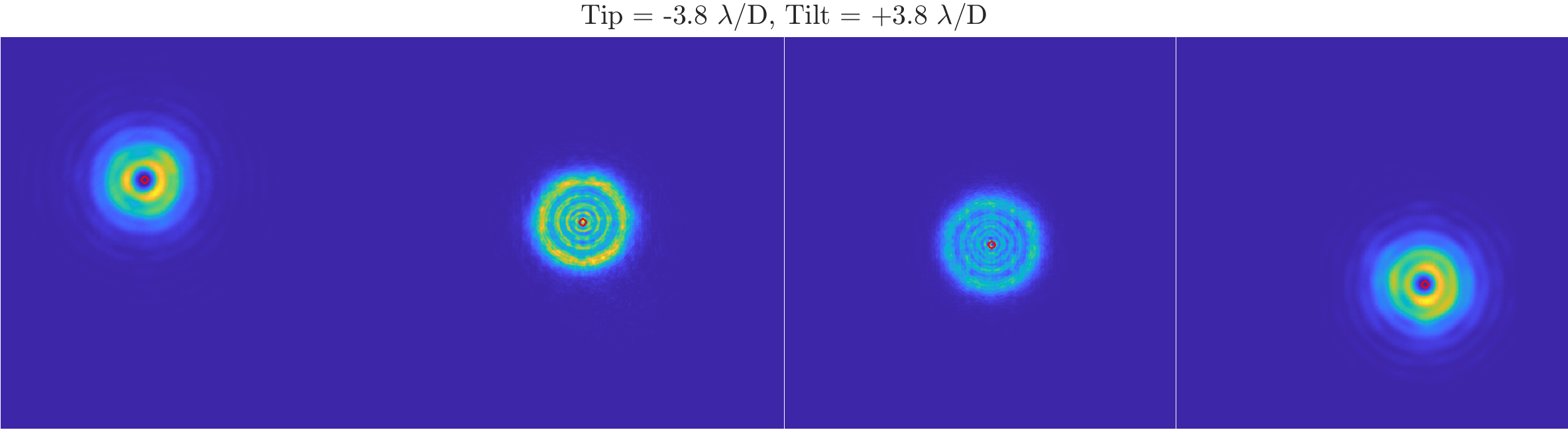}
    \caption{Tip/tilt calibration images generated at the measurement planes with no aberrations (top) and a known amount of injected tip/tilt (bottom). WA centroid estimation is shown as a red circle in the bottom figure.}
    \label{fig:calib}
\end{figure}

For tip/tilt calibration, a two-step Hough–Canny approach is used to determine the center of an unaberrated beam in each imaging plane. First, a Canny edge detector identifies the beam’s boundary \cite{Canny1986}. Then, a Circular Hough Transform (CHT) fits a circle to the detected edge and extracts the circle’s estimated center as the reference centroid \cite{Duda1972}. This method assumes the calibration beam is symmetric and circular, a condition verified in Figure \ref{fig:calib}.

\subsection{Closed-Loop Tip/Tilt Control} \label{sec:clc}

Implementing tip/tilt correction in a closed-loop environment is critical for operation. A standard PI (proportional–integral) control scheme is used to drive the FSM, enabling correction of tip/tilt offsets while minimizing the risk of overcorrection and long-term drift. The closed-loop approach removes the observer from the control process, allowing the system to operate autonomously, whereas previous corrections and analysis were performed offline in post-processing. For the closed-loop data presented here, the system follows a focal-plane correction strategy for the nlCWFS similar to that demonstrated by Crepp et al. (2025) \cite{Crepp2025}. This configuration uses a 200 mm focal length lens with measurement planes located at $z = {150,\ 160,\ 170,\ 180}$ mm, measured from the lens.

For the preliminary experiments described in Section~\ref{sec:cl_results}, the PI coefficients are heuristically set to $P=0.5$ and $I=1.0$. These values were chosen to illustrate loop stability and convergence, rather than to achieve optimal response speed or bandwidth. No gain normalization was applied. Tip and tilt offsets are estimated using the same WA centroiding method described in Section~\ref{sec:centr_calib}. Images are acquired every $0.04$ seconds. An initial tip/tilt offset of $\theta_{\mathrm{Tip}}=-500$" and $\theta_{\mathrm{Tilt}}=500$" ($\pm 1.92\,\lambda/D$) is applied to the FSM outside the control loop to serve as the initial error to be corrected. The control loop is implemented in C++ to support real-time execution. 

While the loop operates at $25$ Hz in this configuration, it is primarily intended to address low-frequency tip/tilt effects such as platform jitter or mechanical drift. Extension to higher-frequency correction for atmospheric turbulence would require further optimization of the centroiding and control system.

\section{RESULTS}\label{sec:results}

\subsection{Tip/Tilt Retrieval (Injected Offsets)} \label{sec:known_res}

\begin{figure}
    \centering
    \includegraphics[width=0.9\linewidth]{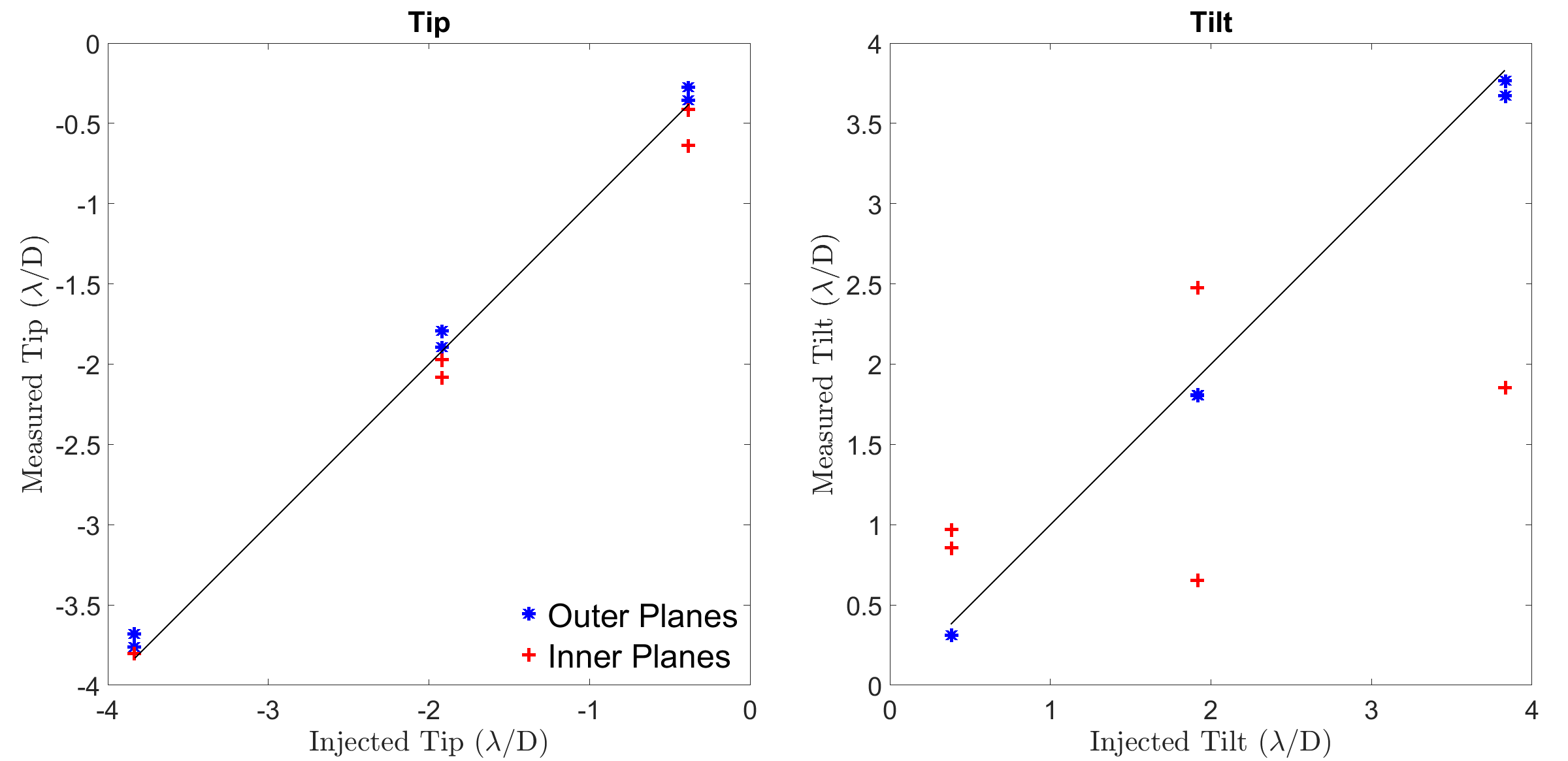}
    \includegraphics[width=0.88\linewidth]{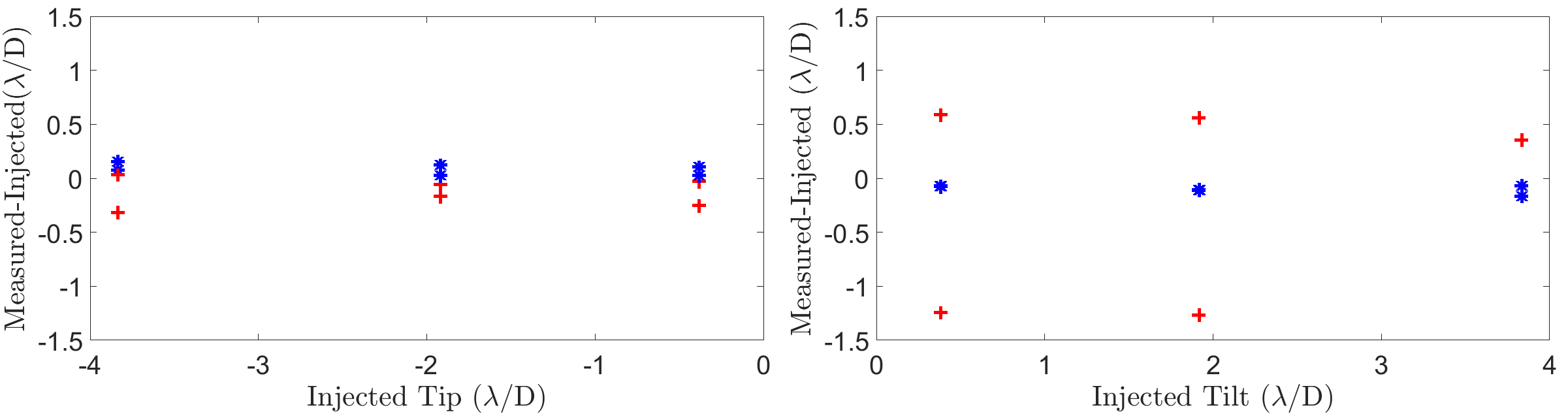}
    \caption{Measured versus injected tip (top left) and tilt (top right) using the WA method and their residuals (bottom). The outer measurement planes (blue) offer more consistent and reliable tip/tilt estimates due to having a larger geometric lever arm. The inner measurement planes (red) require careful calibration and are otherwise less reliable.}
    \label{fig:known_rmse}
\end{figure}

Known amounts of tip and tilt were injected into the optical system using the FSM. Three test cases were evaluated, with tip offsets of $\theta_{\mathrm{Tip}} = \{-3.84,\ -1.92,\ -0.38\}\,\lambda/D$, and corresponding tilt offsets of $\theta_{\mathrm{Tilt}} = \{+3.84,\ +1.92,\ +0.38\}\,\lambda/D$. These values were selected randomly to evaluate the accuracy of the retrieval procedure across a moderate dynamic range, while remaining within the operational limits of the FSM and camera. Figure~\ref{fig:calib} shows the measurement planes and corresponding WA centroid positions for the zero tip/tilt case and the $\pm 3.84 \, \lambda/D$ offset condition. The accuracy of the retrieved values is shown in Figure~\ref{fig:known_rmse}, which plots the measured versus injected phase along with the residuals. For an otherwise unaberrated beam, the WA algorithm, when averaging the results from the outer planes, is able to estimate the tip and tilt to within $\pm0.1 \,\lambda/D$ on average across the three cases. In contrast, the inner planes exhibit a wider spread in their estimates ($\pm0.57\;\lambda/D$) and are less reliable for tip/tilt retrieval.

Several sources of statistical and systematic errors may be considered as contributing to uncertainties in measured tip/tilt. One potential contributor is the FSM’s intrinsic accuracy, which is specified as approximately 2\% of its full range (equivalent to $\pm 0.41\,\lambda/D$). Meanwhile, the device’s angular resolution is much finer, around $2 \, \mu$rad or $0.0016 \, \lambda/D$ \cite{OIM5001}. The FSM is precise and its accuracy is diffraction-limited over small movements. Given the several $\lambda/D$ angular deviations introduced in our experiment, FSM electro-mechanical functionality is therefore not a dominant contributor to overall performance.

Another source of uncertainty is evident in the reduced accuracy of the inner-plane measurements compared with outer-plane measurements. The subtended angle, $\theta_N$, for a region of $N$ pixels at a distance $z$ from the pupil is
\begin{equation} \label{eqn:pix_ang}
    \theta_N = \arctan\left(\frac{N \; \delta_{\rm pixel}}{z}\right),
\end{equation}
where $\delta_{\rm pixel}$ is the pixel plate-scale. Since the inner-planes are positioned closer to the pupil ($0.273\,\lambda/D$ per pixel) than the outer-planes ($0.055\,\lambda/D$ per pixel), small angular shifts are more difficult to measure, making near-field tip/tilt estimates sensitive to WA centroiding accuracy (statistical) and optical mis-alignments (systematic). These results support the use of outer-plane data alone for tip/tilt estimation due to their finer angular resolution and greater sensitivity to low spatial frequency modes.

\subsection{Tip/Tilt Compensation and Reconstruction} \label{sec:closed}

To evaluate the impact of tip/tilt detection and correction on wavefront reconstruction, we emulate closed-loop operation using image-domain adjustments. Based on the calculated angular offsets from the outer measurement planes, a corrective pixel shift is applied to each image using Equation~\ref{eqn:pix_ang} by solving for $N$. After applying the correction, phase reconstructions are performed using a modified Gerchberg-Saxton (GS) algorithm with up to five iterations. See Letchev et al. 2024 \cite{Letchev2024} for a more in-depth discussion of the reconstruction algorithm.

\subsubsection{Reconstruction Without Aberrations (Injected Offsets)} \label{sec:noaber_recon}

\begin{figure}
    \centering
    \includegraphics[width=0.65\linewidth]{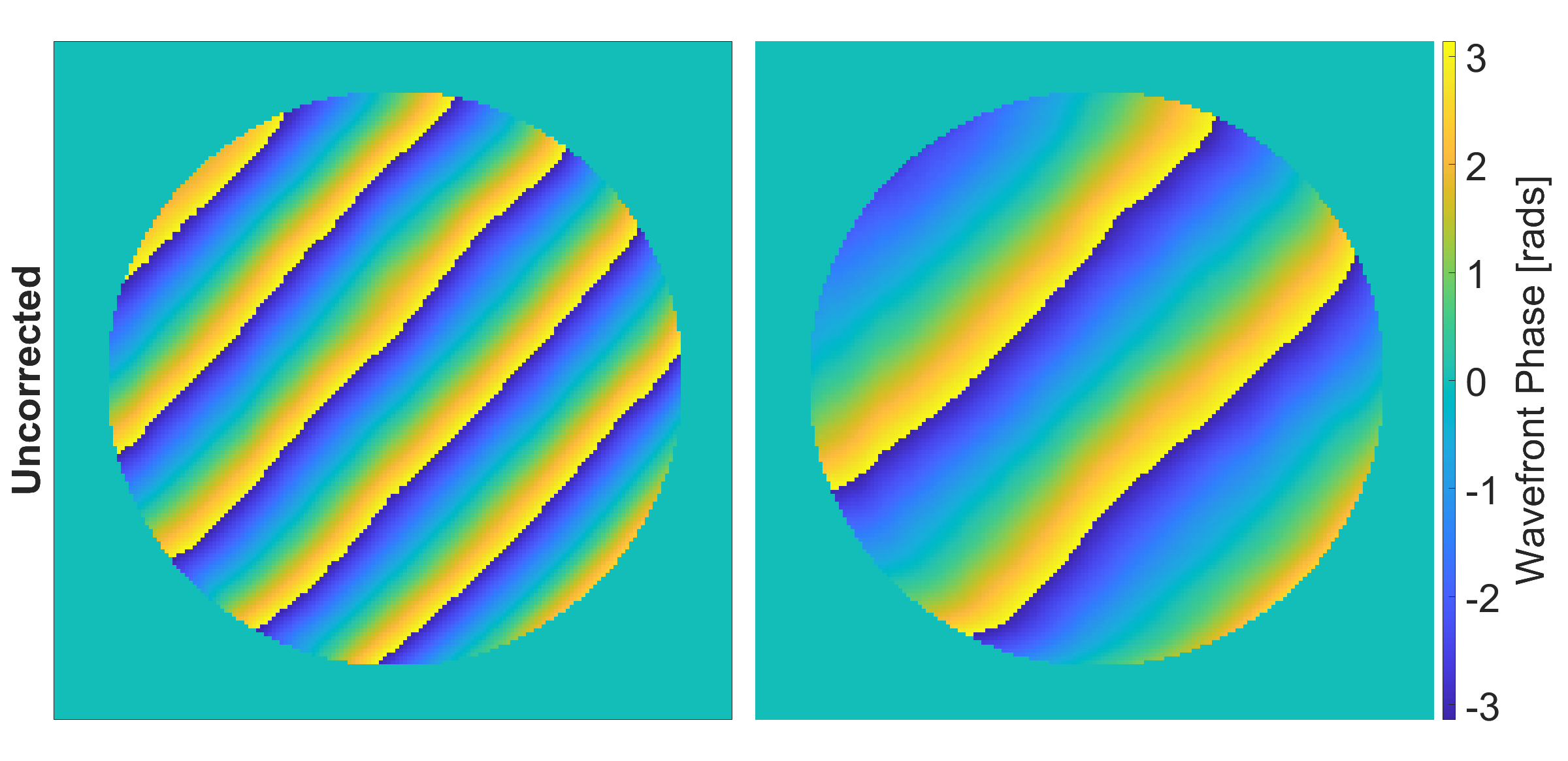}
    \includegraphics[width=0.65\linewidth]{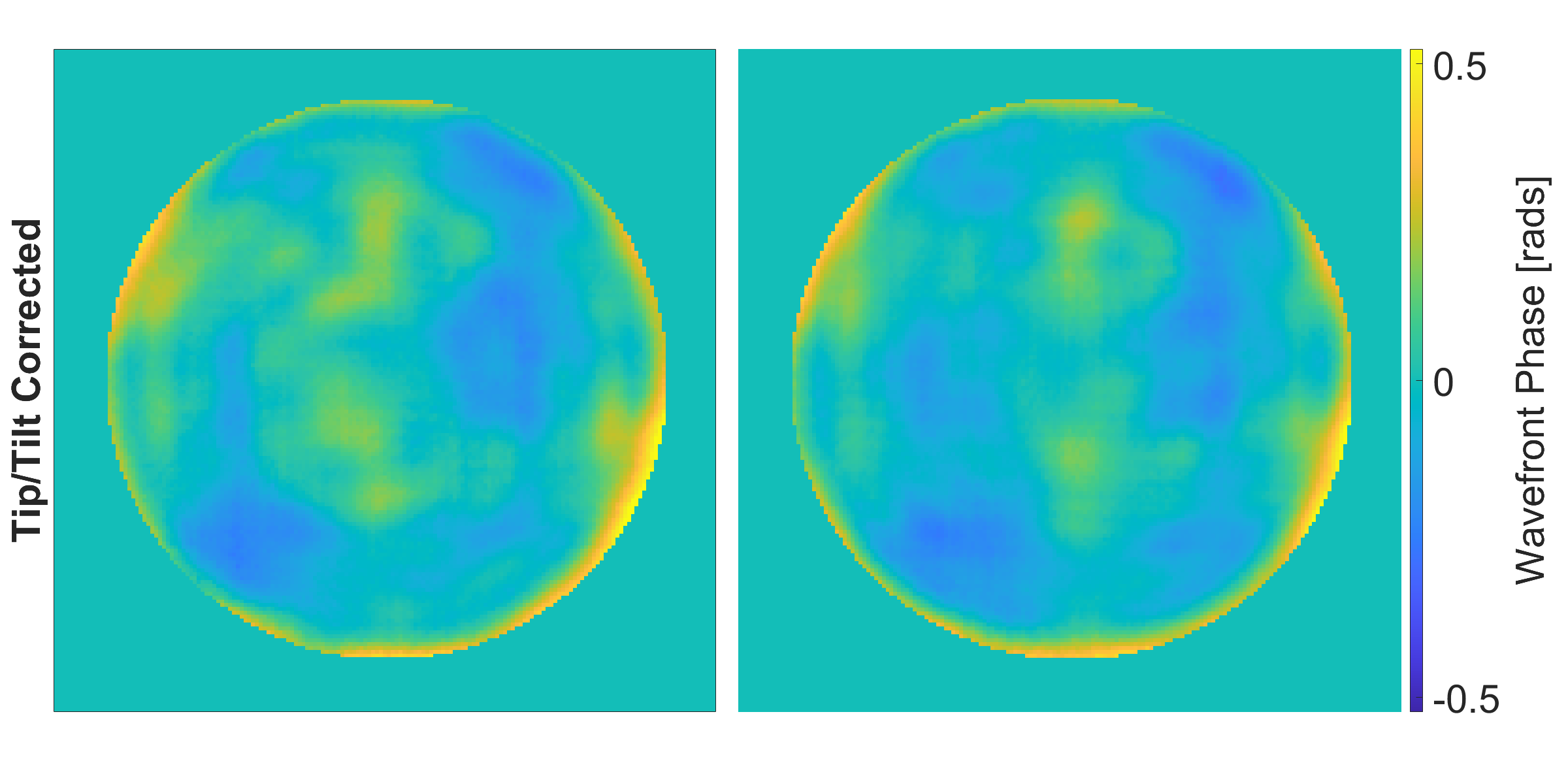}
    \caption{Wrapped phase reconstructions for known tip/tilt cases without (top) and with (bottom) tip/tilt correction. On the left the injected tip/tilt was $-3.84\;\lambda/D$ and $+3.84\;\lambda/D$  while on the right it was $-1.92\;\lambda/D$ and $+1.92\;\lambda/D$. Tip/tilt correction improves wavefront reconstruction by reducing the amplitude of pointing modes, allowing previously hidden features to be more easily observed.}
    \label{fig:known_recon}
\end{figure}

To establish a baseline, the tip/tilt correction procedure was first applied to the non-aberrated datasets. Figure~\ref{fig:known_recon} shows the phase reconstructions before (top row) and after (bottom row) applying the pixel-space correction derived from the outer-plane estimates. As expected, in the absence of additional aberrations, the correction yields a significantly flattened wavefront, confirming that the tip/tilt estimation and compensation method is functioning as intended.

\subsubsection{Reconstruction with Aberrations (No Injected Offsets)} \label{sec:aberr_recon}

\begin{figure}
    \centering
    \includegraphics[width=.75\linewidth]{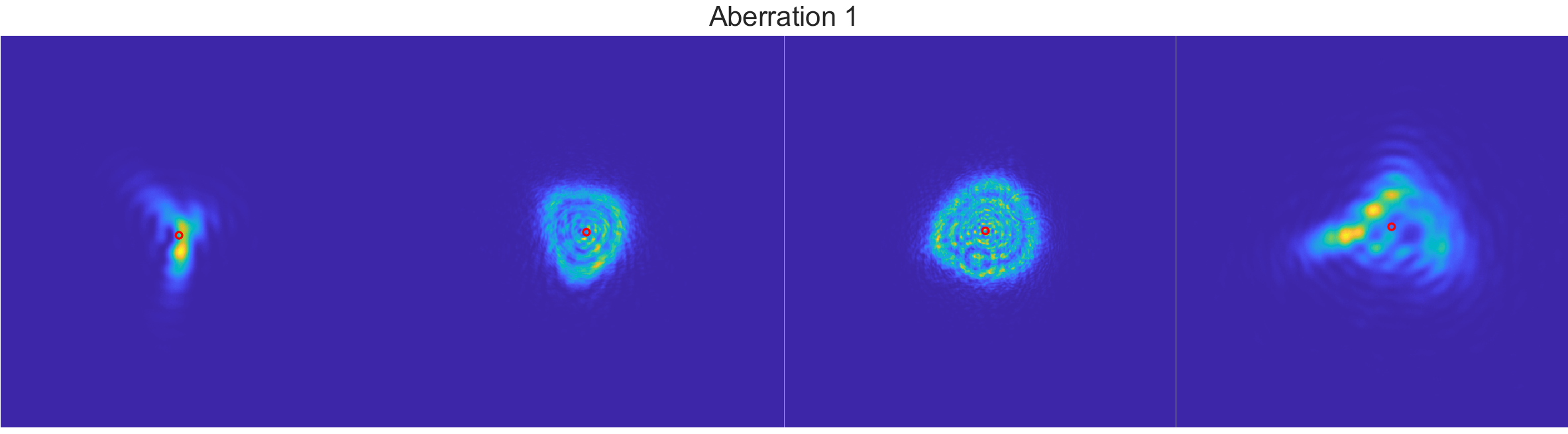}
    \includegraphics[width=.75\linewidth]{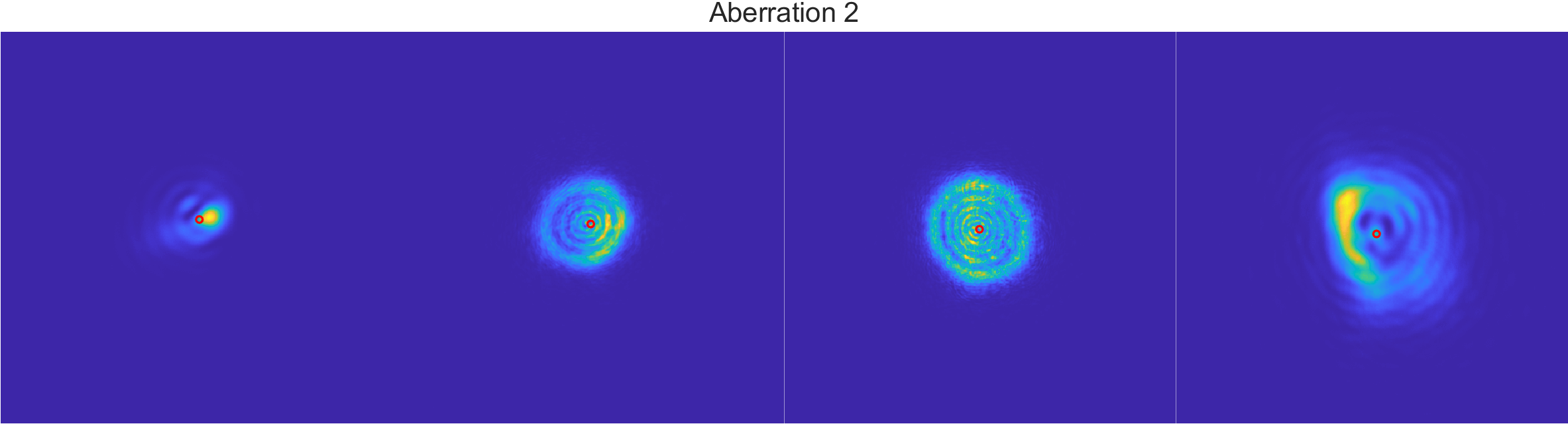}
    \includegraphics[width=.75\linewidth]{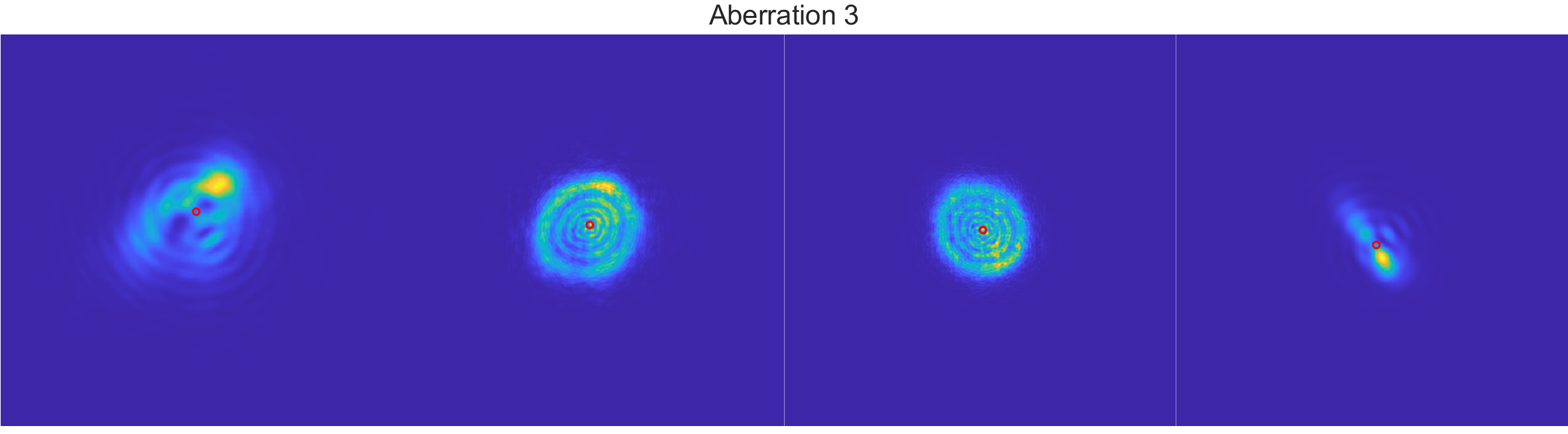}
    \caption{Measurement planes and centroiding estimates (red circles) for Aberration 1, Aberration 2, and Aberration 3.}
    \label{fig:aberr_planes}
\end{figure}

\begin{figure}
    \centering
    \includegraphics[width=0.8\linewidth]{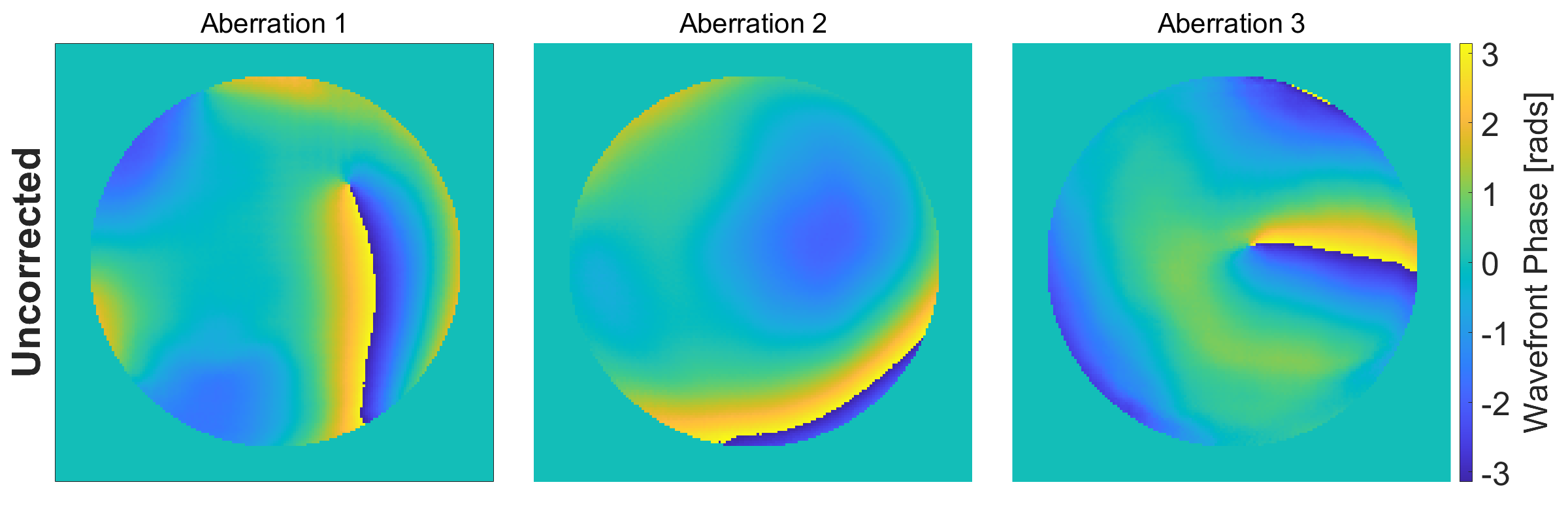}
    \includegraphics[width=0.8\linewidth]{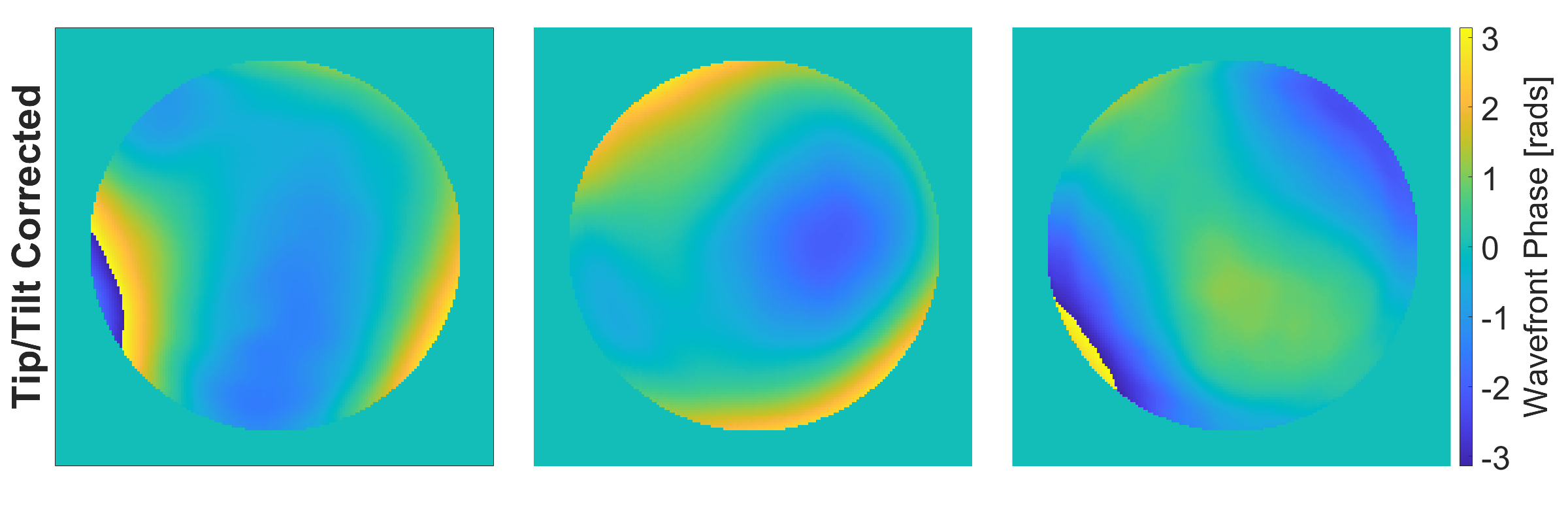}
    \caption{Wrapped phase reconstructions for three different aberrations without (top) and with (bottom) tip/tilt correction. Correction improves reconstruction by reducing the impact of low order modes. For instance, the ``branch cuts" in Aberrations 1 and 3 are no longer present and overall the reconstructions are smoother.}
    \label{fig:aberr_recon}
\end{figure}

To evaluate the effectiveness of tip/tilt correction under more realistic conditions, an aberrator plate was placed near the pupil plane (just upstream of the “telescope”) to induce phase distortions while minimizing scintillation. Figure~\ref{fig:aberr_planes} shows the resulting measurement planes for three different aberration cases. Unlike the known tip/tilt trials, no additional mis-pointing was injected; misalignments arise purely from the aberrated wavefront. A WA centroiding analysis was applied to each spot image, and the resulting centroids are overlaid as red circles in the figure.

\begin{figure}
    \centering
    \includegraphics[width=0.75\linewidth]{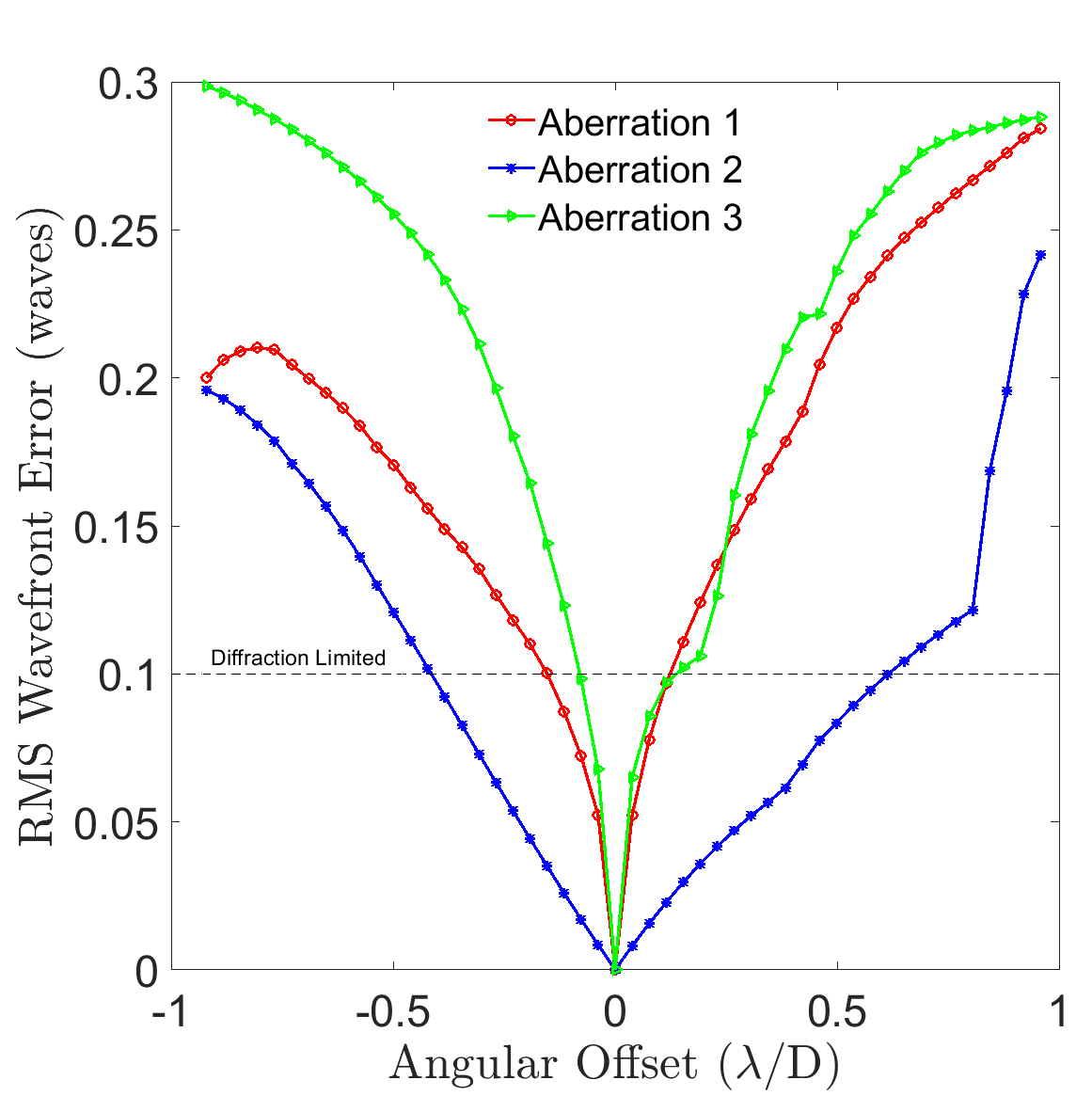}
    \caption{RMS Wavefront Error of phase reconstructions as a function of tip/tilt error. Small offsets in tip/tilt can cascade into large uncertainties in the reconstruction of higher order modes.}
     \label{fig:uplot}
\end{figure}

The corresponding phase reconstructions, before and after tip/tilt correction, are shown in Figure~\ref{fig:aberr_recon}. Improvements are evident visually in the corrected results, particularly for Aberrations 1 and 3, where the reconstructions no longer exhibit branch cuts and the wavefronts appear smoother. These changes suggest that the tip/tilt estimation and correction procedure is performing effectively. This conclusion is further supported by Figure~\ref{fig:uplot}, which shows the root mean square (RMS) wavefront reconstruction error as a function of applied tip/tilt offset. RMS values were computed by applying additional tip and tilt offsets in post-processing to the aberrated datasets in Figure~\ref{fig:aberr_recon}, reconstructing the phase, and then comparing each result to the corresponding tip/tilt-corrected reconstruction. For each aberration, small errors in tip/tilt estimation can lead to large reconstruction inaccuracies. To achieve diffraction-limited performance, the required tip/tilt accuracy is $\approx \pm0.5\,\lambda/D$ for Aberration 2, and $\approx \pm0.2\,\lambda/D$ for Aberrations 1 and 3.

Based on prior simulations \cite{Abbott2025}, the WA method is expected to achieve an accuracy of roughly $\pm 0.3\,\lambda/D$, which is comparable to the requirements estimated from lab data using a single measurement. In practice, integrating WA into a closed-loop AO system that drives tip/tilt modes toward null provides a path forward for ensuring reliable wavefront reconstructions. In other words, as tip/tilt correction improves, wavefront reconstructions improve, making the measurement plane images more symmetric, which in turn allows WA to achieve improved centroid accuracy (approximately $\pm 0.1\,\lambda/D$ in the unaberrated case; Figure~\ref{fig:known_rmse}, Section \ref{sec:known_res}), establishing a positive feedback loop for the nlCWFS.

\subsubsection{Closed-Loop Experimental Results (Injected Offsets)} \label{sec:cl_results}

\begin{figure}
    \centering
    \includegraphics[width=0.75\linewidth]{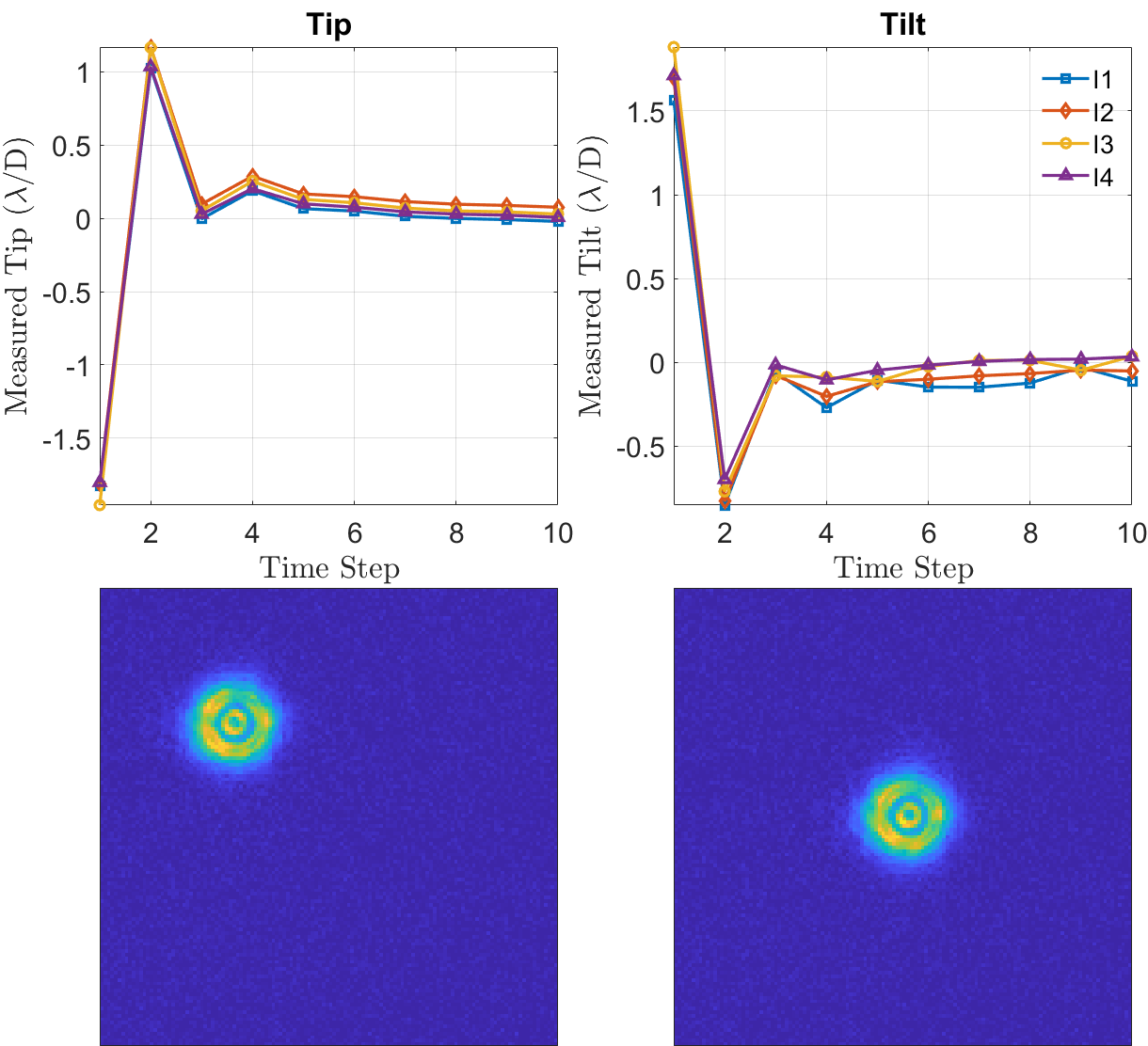}
    \caption{Measured angular offset in tip and tilt during active closed loop control (top). Starting and ending image for the fourth measurement plane (I4) after tip/tilt correction (bottom).}
     \label{fig:lab_norm_ttm}
\end{figure}

Building on the structure and approach described above, a closed-loop tip/tilt control sequence was implemented using the coefficients and design from Section~\ref{sec:clc}. Figure~\ref{fig:lab_norm_ttm} shows the results of this control loop applied to an unaberrated beam (excluding the initial offset introduced for correction). Within 3–5 time steps, the FSM successfully drives the imaging planes toward null. Further tuning of the PID coefficients and control system are necessary depending on the amount of turbulence and application to achieve the desired and necessary speeds.

\begin{figure}
    \centering
    \includegraphics[width=0.75\linewidth]{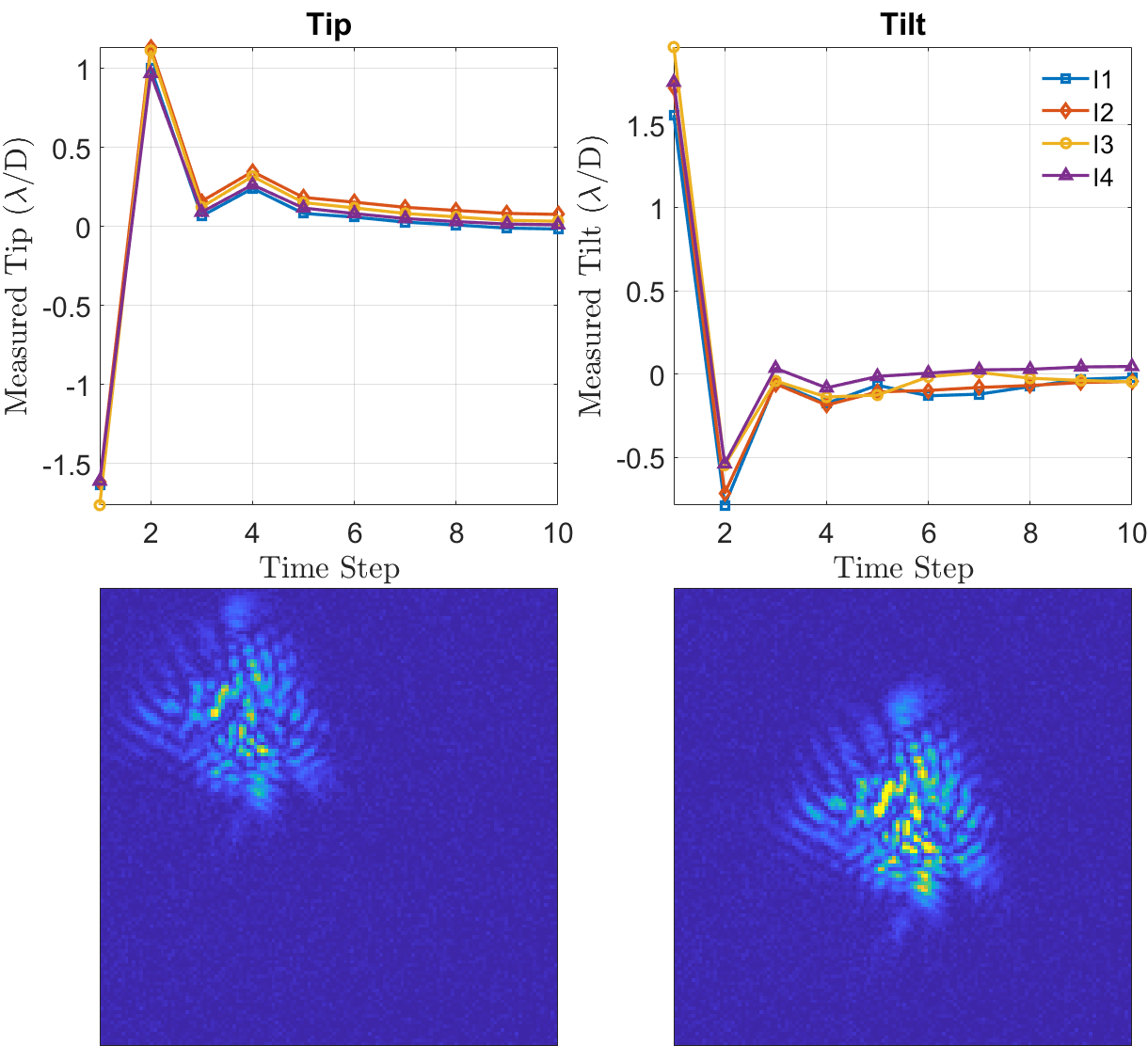}
    \caption{Measured angular offset in tip and tilt during active closed loop control with an aberration (top). Starting and ending image for the fourth measurement plane (I4) after tip/tilt correction (bottom).}
     \label{fig:lab_aberr_ttm}
\end{figure}

The same control setup, with an additional aberration applied to the beam, is shown in Figure~\ref{fig:lab_aberr_ttm}. Similar performance is observed where within a few iterations the FSM again centers the imaging planes near the array center. These results demonstrate, for the first time, the ability to compensate for tip/tilt errors using only a FSM and the intensity data already acquired for wavefront reconstruction in hardware with the nlCWFS. A forthcoming article will detail the implementation and results of operating this closed-loop tip/tilt control in tandem with closed-loop high-order correction via a deformable mirror.

\section{Summary and Concluding Remarks} \label{sec:summary}

This work shows that tip/tilt can be extracted concurrently with high-order phase retrieval using only the internal optics of the nlCWFS, without relying on quad-cells or additional cameras. Building on previous numerical work, which showed tip/tilt reconstruction accuracy of $\pm 0.3\,\lambda/D$, we have validated theoretical approaches in a laboratory setting, achieving comparable results of $\pm0.1\,\lambda/D$ for an unaberrated beam and better than $\sim\pm0.5 \;\lambda/D$ in the presence of aberrations. The experiments corroborate that the outer planes provide the best estimate of tip/tilt due to the longer lever arm. Across different experiments, the simple WA centroiding method proved to be an accurate and reliable estimator for detecting tip/tilt offsets.

These results support a vision of a closed-loop AO system in which tip/tilt compensation is integrated directly into the nlCWFS reconstruction pipeline. Such a feedback loop would iteratively improve both wavefront correction and centroid estimation: tip/tilt correction by the FSM flattens the wavefront, leading to more accurate reconstructions, which in turn enhance centroid estimates and further reduce pointing jitter. In this work, we have demonstrated closed-loop FSM control for tip/tilt correction. Future efforts will focus on (i) optimizing the closed-loop control sequence to achieve faster operation; (ii) quantitatively validating the positive feedback loop through a more detailed analysis of wavefront error before and after tip/tilt correction; and (iii) extending this capability to operate in tandem with deformable mirror--based high-order correction.

\section*{Data Availability} \label{data_avail}
The data used to generate figures for this article are available upon request from the authors.  

\section*{Disclosures}
The authors declare there are no financial interests, commercial affiliations, or other potential conflicts of interest that have influenced the objectivity of this research or the writing of this paper. 

\acknowledgments 

This research was supported in part by the Air Force Office of Scientific Research (AFOSR) grant number FA9550-22-1-0435 and the Joint Directed Energy Transfer Office (JDETO). A preliminary subset of this work was previously included in SPIE Proceedings, Abbott et al. 2025 \cite{Abbott2025b} and uploaded to arXiv (arXiv:2508.09256).

The authors used the AI tool ChatGPT (OpenAI, GPT-4) to assist with grammar, spelling, and language clarity during manuscript preparation. No content generation, data analysis, or figure creation was performed using AI tools.

\bibliography{abbott} 

@article{Abbott2025,
author = {Caleb G. Abbott and Justin R. Crepp and Stanimir O. Letchev and Connor M. Smith},
title = {{Centroiding and extraction of tip/tilt information from nonlinear curvature wavefront sensor measurements}},
volume = {11},
journal = {Journal of Astronomical Telescopes, Instruments, and Systems},
number = {1},
publisher = {SPIE},
pages = {019001},
year = {2025},
doi = {10.1117/1.JATIS.11.1.019001},
URL = {https://doi.org/10.1117/1.JATIS.11.1.019001}
}

@inproceedings{Abbott2025b,
author = {Caleb G. Abbott and Justin R. Crepp and Brian Sands},
title = {{Jitter sensing and control for multiplane phase retrieval}},
volume = {13619},
booktitle = {Unconventional Imaging, Sensing, and Adaptive Optics 2025},
editor = {Jean J. Dolne and Santasri R. Bose-Pillai and Matthew Kalensky},
organization = {International Society for Optics and Photonics},
publisher = {SPIE},
pages = {136191U},
year = {2025},
doi = {10.1117/12.3063658},
URL = {https://doi.org/10.1117/12.3063658}
}

@inproceedings{Ahn2023b,
author = {K. Ahn and O. Guyon and J. Lozi and S. Vievard and V. Deo and M. Lallement and J. C. Bragg},
title = {{A non-linear curvature wavefront sensor for the Subaru telescope’s AO3k system}},
volume = {12680},
booktitle = {Techniques and Instrumentation for Detection of Exoplanets XI},
editor = {Garreth J. Ruane},
organization = {International Society for Optics and Photonics},
publisher = {SPIE},
pages = {126800B},
year = {2023},
doi = {10.1117/12.2676667},
URL = {https://doi.org/10.1117/12.2676667}
}

@inproceedings{Avicola1998,
author = {Kenneth Avicola and James A. Watson and Barton V. Beeman and Thomas C. Kuklo and John R. Taylor},
title = {{Design and performance of the tip-tilt subsystem for the Keck II telescope adaptive optics system}},
volume = {3353},
booktitle = {Adaptive Optical System Technologies},
editor = {Domenico Bonaccini and Robert K. Tyson},
organization = {International Society for Optics and Photonics},
publisher = {SPIE},
pages = {628 -- 637},
year = {1998},
doi = {10.1117/12.321629},
URL = {https://doi.org/10.1117/12.321629}
}

@ARTICLE{Bechter2020,
       author = {{Bechter}, Andrew J. and {Crass}, Jonathan and {Tesch}, Jonathan and {Crepp}, Justin R. and {Bechter}, Eric B.},
        title = "{Characterization of Single-mode Fiber Coupling at the Large Binocular Telescope}",
      journal = {PASP},
         year = 2020,
        month = jan,
       volume = {132},
       number = {1007},
          eid = {015001},
        pages = {015001},
          doi = {10.1088/1538-3873/ab42cb},
archivePrefix = {arXiv},
       eprint = {2005.10737},
 primaryClass = {astro-ph.IM},
       adsurl = {https://ui.adsabs.harvard.edu/abs/2020PASP..132a5001B}
}

@ARTICLE{Crepp2014,
       author = {{Crepp}, Justin R.},
        title = "{Improving planet-finding spectrometers}",
      journal = {Science},
         year = 2014,
        month = nov,
       volume = {346},
       number = {6211},
        pages = {809-810},
          doi = {10.1126/science.1262071},
archivePrefix = {arXiv},
       eprint = {1412.2992},
 primaryClass = {astro-ph.IM},
       adsurl = {https://ui.adsabs.harvard.edu/abs/2014Sci...346..809C}
}

@inproceedings{Crepp2025,
author = {Justin R. Crepp and Caleb G. Abbott and James Smous and Matthew Engstrom and Brian Sands},
title = {{Phase retrieval using multiplane intermediate-focal coherent diffractive wavefront sensing}},
volume = {13619},
booktitle = {Unconventional Imaging, Sensing, and Adaptive Optics 2025},
editor = {Jean J. Dolne and Santasri R. Bose-Pillai and Matthew Kalensky},
organization = {International Society for Optics and Photonics},
publisher = {SPIE},
pages = {136191V},
year = {2025},
doi = {10.1117/12.3063659},
URL = {https://doi.org/10.1117/12.3063659}
}

@article{Canny1986,
  author = {Canny, John},
  title = {A computational approach to edge detection},
  journal = {IEEE Transactions on Pattern Analysis and Machine Intelligence},
  volume = {PAMI-8},
  number = {6},
  pages = {679--698},
  year = {1986},
  doi = {10.1109/TPAMI.1986.4767851}
}

@phdthesis{Crass2014b,
author = {Crass, Jonathan},
year = {2014},
month = {07},
pages = {},
title = {The Adaptive Optics Lucky Imager: combining adaptive optics and lucky imaging},
school = {University of Cambridge}
}

@article{Crepp2020,
author = {Justin R. Crepp and Stanimir O. Letchev and Sam J. Potier and Joshua H. Follansbee and Nicholas T. Tusay},
journal = {Opt. Express},
number = {25},
pages = {37721--37733},
publisher = {Optica Publishing Group},
title = {Measuring phase errors in the presence of scintillation},
volume = {28},
month = {12},
year = {2020},
url = {https://opg.optica.org/oe/abstract.cfm?URI=oe-28-25-37721},
doi = {10.1364/OE.408825},
}

@article{Duda1972,
  author = {Duda, Richard O. and Hart, Peter E.},
  title = {Use of the Hough transformation to detect lines and curves in pictures},
  journal = {Communications of the ACM},
  volume = {15},
  number = {1},
  pages = {11--15},
  year = {1972},
  doi = {10.1145/361237.361242}
}

@article{Guyon2010,
 ISSN = {00046280, 15383873},
 URL = {http://www.jstor.org/stable/10.1086/649646},
 author = {Olivier Guyon},
 journal = {Publications of the Astronomical Society of the Pacific},
 number = {887},
 pages = {49--62},
 publisher = {[The University of Chicago Press, Astronomical Society of the Pacific]},
 title = {High Sensitivity Wavefront Sensing with a Nonlinear Curvature Wavefront Sensor},
 urldate = {2023-12-14},
 volume = {122},
 year = {2010}
}

@inproceedings{Huerta2025,
author = {D. Angelica Huerta and Justin R. Crepp and Caleb G. Abott and Brian Joseph},
title = {{Comparison of phase-unwrapping methods for adaptive optics wavefront sensing}},
volume = {13619},
booktitle = {Unconventional Imaging, Sensing, and Adaptive Optics 2025},
editor = {Jean J. Dolne and Santasri R. Bose-Pillai and Matthew Kalensky},
organization = {International Society for Optics and Photonics},
publisher = {SPIE},
pages = {136191X},
year = {2025},
doi = {10.1117/12.3063929},
URL = {https://doi.org/10.1117/12.3063929}
}

@article{Kong2017,
author = {Fanpeng Kong and Manuel Cegarra Polo and Andrew Lambert},
journal = {Appl. Opt.},
number = {23},
pages = {6466--6475},
publisher = {Optica Publishing Group},
title = {Centroid estimation for a Shack--Hartmann wavefront sensor based on stream processing},
volume = {56},
month = {Aug},
year = {2017},
url = {https://opg.optica.org/ao/abstract.cfm?URI=ao-56-23-6466},
doi = {10.1364/AO.56.006466},
}

@INPROCEEDINGS{Letchev2022,
       author = {{Letchev}, Stanimir and {Crass}, Jonathan and {Crepp}, Justin R. and {Potier}, Sam},
        title = "{Spatial frequency response and sensitivity of the nonlinear curvature wavefront sensor}",
    booktitle = {Adaptive Optics Systems VIII},
         year = 2022,
       editor = {{Schreiber}, Laura and {Schmidt}, Dirk and {Vernet}, Elise},
       series = {Society of Photo-Optical Instrumentation Engineers (SPIE) Conference Series},
       volume = {12185},
        month = aug,
          eid = {121858H},
        pages = {121858H},
          doi = {10.1117/12.2632449},
archivePrefix = {arXiv},
       eprint = {2209.00071},
 primaryClass = {astro-ph.IM},
       adsurl = {https://ui.adsabs.harvard.edu/abs/2022SPIE12185E..8HL}
}

@article{Letchev2023,
author = {Letchev, Stanimir and Crass, Jonathan and Crepp, Justin},
year = {2023},
month = {10},
pages = {},
title = {Assessing phase reconstruction accuracy for different nonlinear curvature wavefront sensor configurations},
volume = {9},
journal = {Journal of Astronomical Telescopes, Instruments, and Systems},
doi = {10.1117/1.JATIS.9.4.049001}
}

@phdthesis{Letchev2024,
author = "Stanimir  Letchev",
title = "{Advancing the Technology Readiness of the Nonlinear Curvature Wavefront Sensor through Modeling and Laboratory Investigation}",
year = "2024",
month = "5",
url = "https://curate.nd.edu/articles/dataset/Advancing_the_Technology_Readiness_of_the_Nonlinear_Curvature_Wavefront_Sensor_through_Modeling_and_Laboratory_Investigation/25607793",
doi = "10.7274/25607793.v1",
school = "University of Notre Dame"
}

@inproceedings{Mateen2011,
author = {Mala Mateen and Olivier Guyon and Jos{\'e} Sasi{\'a}n and Vincent Garrel and Michael Hart},
title = {{A non-linear curvature wavefront sensor reconstruction speed and the broadband design}},
volume = {8149},
booktitle = {Astronomical Adaptive Optics Systems and Applications IV},
editor = {Robert K. Tyson and Michael Hart},
organization = {International Society for Optics and Photonics},
publisher = {SPIE},
pages = {814909},
year = {2011},
doi = {10.1117/12.894311},
URL = {https://doi.org/10.1117/12.894311}
}

@article{Nicolle2004,
author = {M. Nicolle and T. Fusco and G. Rousset and V. Michau},
journal = {Opt. Lett.},
number = {23},
pages = {2743--2745},
publisher = {Optica Publishing Group},
title = {Improvement of Shack--Hartmann wave-front sensor measurement for extreme adaptive optics},
volume = {29},
month = {Dec},
year = {2004},
url = {https://opg.optica.org/ol/abstract.cfm?URI=ol-29-23-2743},
doi = {10.1364/OL.29.002743},
}

@misc{OIM5001,
  author       = {{Optics In Motion LLC}},
  title        = {OIM5001 Small One-Inch Fast Steering Mirror: Specification Sheet},
  year         = {2013},
  note         = {Technical datasheet},
}

@ARTICLE{Potier2023,
       author = {{Potier}, Sam and {Crepp}, Justin and {Letchev}, Stanimir},
        title = "{Developing an error budget for the nonlinear curvature wavefront sensor}",
      journal = {Journal of Astronomical Telescopes, Instruments, and Systems},
         year = 2023,
        month = oct,
       volume = {9},
          eid = {049004},
        pages = {049004},
          doi = {10.1117/1.JATIS.9.4.049004},
archivePrefix = {arXiv},
       eprint = {2312.01465},
 primaryClass = {astro-ph.IM},
       adsurl = {https://ui.adsabs.harvard.edu/abs/2023JATIS...9d9004P}
}

@ARTICLE{Sarazin1990,
       author = {{Sarazin}, M. and {Roddier}, F.},
       title = "{The ESO differential image motion monitor}",
      journal = {Astronomy and Astrophysics},
         year = 1990,
        month = jan,
       volume = {227},
       number = {1},
        pages = {294-300},
       adsurl = {https://ui.adsabs.harvard.edu/abs/1990A&A...227..294S}
}

@article{Thomas2006,
    author = {Thomas, S. and Fusco, T. and Tokovinin, A. and Nicolle, M. and Michau, V. and Rousset, G.},
    title = {Comparison of centroid computation algorithms in a Shack–Hartmann sensor},
    journal = {Monthly Notices of the Royal Astronomical Society},
    volume = {371},
    number = {1},
    pages = {323-336},
    year = {2006},
    month = {08},
    issn = {0035-8711},
    doi = {10.1111/j.1365-2966.2006.10661.x},
    url = {https://doi.org/10.1111/j.1365-2966.2006.10661.x},
    eprint = {https://academic.oup.com/mnras/article-pdf/371/1/323/3029534/mnras0371-0323.pdf},
}

@article{Tokovinin2007,
    author = {Tokovinin, A. and Kornilov, V.},
    title = {Accurate seeing measurements with MASS and DIMM},
    journal = {Monthly Notices of the Royal Astronomical Society},
    volume = {381},
    number = {3},
    pages = {1179-1189},
    year = {2007},
    month = {10},
    issn = {0035-8711},
    doi = {10.1111/j.1365-2966.2007.12307.x},
    url = {https://doi.org/10.1111/j.1365-2966.2007.12307.x},
    eprint = {https://academic.oup.com/mnras/article-pdf/381/3/1179/3656929/mnras0381-1179.pdf},
}

@book{Tyson2022,
  author = {Tyson, Robert K. and Frazier, Benjamin West},
  title = {Principles of Adaptive Optics},
  edition = {5th},
  publisher = {CRC Press},
  year = {2022},
  isbn = {9780367676032},
  doi = {10.1201/9781003140191}
}

@inproceedings{Wizinowich2014,
author = {Peter Wizinowich and Roger Smith and Roberto Biasi and Sylvain Cetre and Richard Dekany and Bruno Femenia-Castella and Jason Fucik and David Hale and Chris Neyman and Dietrich Pescoller and Sam Ragland and Paul Stomski and Mario Andrighettoni and Randy Bartos and Khanh Bui and Andrew Cooper and John Cromer and Marcos van Dam and Michael Hess and Ean James and Jim Lyke and Hector Rodriguez and Thomas Stalcup},
title = {{A near-infrared tip-tilt sensor for the Keck I laser guide star adaptive optics system}},
volume = {9148},
booktitle = {Adaptive Optics Systems IV},
editor = {Enrico Marchetti and Laird M. Close and Jean-Pierre V{\'e}ran},
organization = {International Society for Optics and Photonics},
publisher = {SPIE},
pages = {91482B},
year = {2014},
doi = {10.1117/12.2055279},
URL = {https://doi.org/10.1117/12.2055279}
}
\bibliographystyle{spiebib} 

Caleb G. Abbott is a postdoctoral scholar at the University of Notre Dame. His research has focused on astronomical instrumentation, specifically the advancement of AO technologies. He previously received his PhD degree in astronomy and master's degree in physics from Georgia State University (2022 and 2019 respectively) and bachelor's degrees in astronomy and physics from the University of Michigan (2014).

Justin R. Crepp is an experimental astrophysicist at the University of Notre Dame. He designs and builds instruments for ground-based telescopes. His research focuses on developing technologies related to adaptive optics. Prior to teaching at Notre Dame, he was a postdoctoral scholar at the California Institute of Technology (2008 to 2012). He received his PhD in astronomy from the University of Florida in 2008 and a bachelor’s degree in physics from the Pennsylvania State University in 2003.

Brian Sands is a senior research engineer with the Engineering and Design Core Facility at the University of Notre Dame. His current areas of expertise are in control system design and integration for ground-based astronomical instruments, and development of hardware-based real-time control systems. He received his master's degree in Physics from Miami University in 2005 and a bachelor's degree in Computational Physics from Pennsylvania State University in 2002.

\end{document}